%% file: ecoci.tex
\documentclass[a4paper, 11pt]{article}\usepackage[]{graphicx}\usepackage[dvipsnames]{xcolor}
\makeatletter
\def\maxwidth{ %
  \ifdim\Gin@nat@width>\linewidth
    \linewidth
  \else
    \Gin@nat@width
  \fi
}
\makeatother

\definecolor{fgcolor}{rgb}{0.345, 0.345, 0.345}
\newcommand{\hlnum}[1]{\textcolor[rgb]{0.686,0.059,0.569}{#1}}%
\newcommand{\hlstr}[1]{\textcolor[rgb]{0.192,0.494,0.8}{#1}}%
\newcommand{\hlcom}[1]{\textcolor[rgb]{0.678,0.584,0.686}{\textit{#1}}}%
\newcommand{\hlopt}[1]{\textcolor[rgb]{0,0,0}{#1}}%
\newcommand{\hlstd}[1]{\textcolor[rgb]{0.345,0.345,0.345}{#1}}%
\newcommand{\hlkwb}[1]{\textcolor[rgb]{0.69,0.353,0.396}{#1}}%
\newcommand{\hlkwc}[1]{\textcolor[rgb]{0.333,0.667,0.333}{#1}}%
\newcommand{\hlkwd}[1]{\textcolor[rgb]{0.737,0.353,0.396}{\textbf{#1}}}%

\usepackage{framed}
\makeatletter
\newenvironment{kframe}{%
 \def\at@end@of@kframe{}%
 \ifinner\ifhmode%
  \def\at@end@of@kframe{\end{minipage}}%
  \begin{minipage}{\columnwidth}%
 \fi\fi%
 \def\FrameCommand##1{\hskip\@totalleftmargin \hskip-\fboxsep
 \colorbox{shadecolor}{##1}\hskip-\fboxsep
     \hskip-\linewidth \hskip-\@totalleftmargin \hskip\columnwidth}%
 \MakeFramed {\advance\hsize-\width
   \@totalleftmargin\z@ \linewidth\hsize
   \@setminipage}}%
 {\par\unskip\endMakeFramed%
 \at@end@of@kframe}
\makeatother

\definecolor{shadecolor}{rgb}{.97, .97, .97}
\definecolor{messagecolor}{rgb}{0, 0, 0}
\definecolor{warningcolor}{rgb}{1, 0, 1}
\definecolor{errorcolor}{rgb}{1, 0, 0}
\newenvironment{knitrout}{}{} % an empty environment to be redefined in TeX

\usepackage{alltt}
\usepackage[T1]{fontenc}
\usepackage[utf8]{inputenc}
\usepackage[english]{babel}
\usepackage{graphics}
\usepackage[dvipsnames]{xcolor}
\usepackage{amsmath, amssymb}
\usepackage{doi} % automatic doi-links
\usepackage[round]{natbib} % bibliography
\usepackage{booktabs} % nicer tables
\usepackage[title]{appendix} % better appendices
\usepackage{nameref} % reference appendices with names
\usepackage[onehalfspacing]{setspace} % more space
\usepackage[labelfont=bf,font=small]{caption} % smaller captions

\usepackage{geometry}
\newcommand\longtitle{Evidential Calibration of Confidence Intervals}
\newcommand\shorttitle{Evidential Calibration of Confidence Intervals} % if longtitle too long, change here
\newcommand\subtitle{}
\newcommand\longauthors{Samuel Pawel\textsuperscript{$\star$},
Alexander Ly\textsuperscript{$\dagger\ddagger$},
Eric-Jan Wagenmakers\textsuperscript{$\dagger$}}
\newcommand\shortauthors{S. Pawel, A. Ly, E.-J. Wagenmakers} % if longauthors too long, change here
\newcommand\affiliation{
  $\star$ Department of Biostatistics, University of Zurich \\
  $\dagger$ Psychological Methods, University of Amsterdam \\
  $\ddagger$ Machine Learning Group, Centrum Wiskunde \& Informatica
}
\newcommand\mail{samuel.pawel@uzh.ch}
\title{
  \vspace{-2.5em}
  \textbf{\longtitle} \\
  \subtitle
}
\author{
  \textbf{\longauthors} \\
  \affiliation \\
  E-mail: \href{mailto:\mail}{\mail}
}
\date{June 27, 2023} %don't forget to hard-code date when submitting to arXiv!

\usepackage{hyperref}
\hypersetup{
  bookmarksopen=true,
  breaklinks=true,
  pdftitle={\shorttitle},
  pdfauthor={\shortauthors},
  colorlinks=true,
  linkcolor=blue,
  anchorcolor=black,
  citecolor=blue,
  urlcolor=black,
}

\usepackage{fancyhdr}
\input{defs.tex}
\usepackage{float}
\IfFileExists{upquote.sty}{\usepackage{upquote}}{}
\begin{document}
\maketitle

% %% disclaimer that a preprint
% \begin{center}
%   \vspace{-2em}
%   {\color{red}This is a preprint which has not yet been peer reviewed.}
% \end{center}

%% Abstract
%% -----------------------------------------------------------------------------
\begin{center}
  \begin{minipage}{14cm} {\small
      \rule{\textwidth}{0.5pt} \\
      {\centering \textbf{Abstract} \\
        We present a novel and easy-to-use method for calibrating error-rate
        based confidence intervals to evidence-based support intervals. Support
        intervals are obtained from inverting Bayes factors based on a parameter
        estimate and its standard error. A $k$~support interval can be
        interpreted as ``the observed data are at least $k$ times more likely
        under the included parameter values than under a specified
        alternative''. Support intervals depend on the specification of prior
        distributions for the parameter under the alternative, and we present
        several types that allow different forms of external knowledge to be
        encoded. We also show how prior specification can to some extent be
        avoided by considering a class of prior distributions and then computing
        so-called minimum support intervals which, for a given class of priors,
        have a one-to-one mapping with confidence intervals. We also illustrate
        how the sample size of a future study can be determined based on the
        concept of support. Finally, we show how the bound for the type I error
        rate of Bayes factors leads to a bound for the coverage of support
        intervals. An application to data from a clinical trial illustrates how
        support intervals can lead to inferences that are both intuitive and
        informative.}
      \rule{\textwidth}{0.4pt} }
\end{minipage}
\end{center}

\section{Introduction}
A pervasive problem in data analysis is to draw inferences about unknown
parameters of statistical models. For instance, data analysts are often
interested in identifying a set of parameter values which are relatively
compatible with the observed data. Here we focus on a particular method for
doing so --- the \emph{support set} --- that arguably represents
a %a natural Bayesian and likelihoodist
natural evidential answer to the problem both from a likelihoodist
\citep{Edwards1971, Royall1997, Blume2002} and a Bayesian
\citep{Wagenmakers2020} point of view. In either paradigm, statistical evidence
may be defined via the \emph{Law of Likelihood} \citep{Hacking1965}, that is,
data constitute evidence for one parameter value over an alternative parameter
value if the likelihood of the data under that parameter value is larger than
under the alternative parameter value. The likelihood ratio (or Bayes factor)
measures the strength of evidence, and it plays also a central role in the
construction of support sets, as we will explain in the following.

Let $f(x \given \theta)$ denote the likelihood of the observed data $x$. Let
$\theta$ be an unknown parameter and denote by
\begin{align}
    \BF_{01}(x; \theta_0)
    \label{eq:bf}
    % = \frac{f(x\given\h{0}\colon \theta = \theta_0)}{
    % f(x\given\h{1}\colon \theta \neq \theta_0)}
    = \frac{f(x\given\h{0})}{f(x\given\h{1})}
    = \frac{f(x\given\theta_0)}{\int f(x\given \theta) \,
    f(\theta \given \h{1})\,\text{d}\theta}
\end{align}
the Bayes factor quantifying the strength of evidence which the observed data
$x$ provide for the simple null hypothesis $\h{0} \colon \theta = \theta_0$
relative to a (possibly composite) alternative hypothesis
$\h{1} \colon \theta \neq \theta_0$, with $f(x\given\h{1})$ the marginal
likelihood of $x$ obtained from integrating the likelihood $f(x \given \theta)$
with respect to the prior density of the parameter $f(\theta \given \h{1})$
under the alternative $\h{1}$ % . If a point prior is assigned to $\theta$ under
% $\h{1}$, the Bayes factor reduces to a likelihood ratio
\citep{Jeffreys1961, Kass1995}. For constructing a support interval, one views
the Bayes factor~\eqref{eq:bf} %$\BF_{01}(x; \theta_0)$
as a function of the null value $\theta_0$ for fixed data $x$. A $k$
\emph{support set} for $\theta$ is then given by the set of parameter values for
which the data are $k$ times more likely than under the alternative hypothesis
$\h{1}$ \citep{Wagenmakers2020}, that is,
\begin{align}
    \label{eq:ss}
    \mbox{SI}_k = \big\{\theta_0 : \BF_{01}(x; \theta_0) \geq k \big\}.
\end{align}
The support set thus includes the parameter values for which the observed data
provide statistical evidence of at least level $k$.

\begin{figure*}[!b]
\begin{knitrout}
\definecolor{shadecolor}{rgb}{0.969, 0.969, 0.969}\color{fgcolor}
\includegraphics[width=\maxwidth]{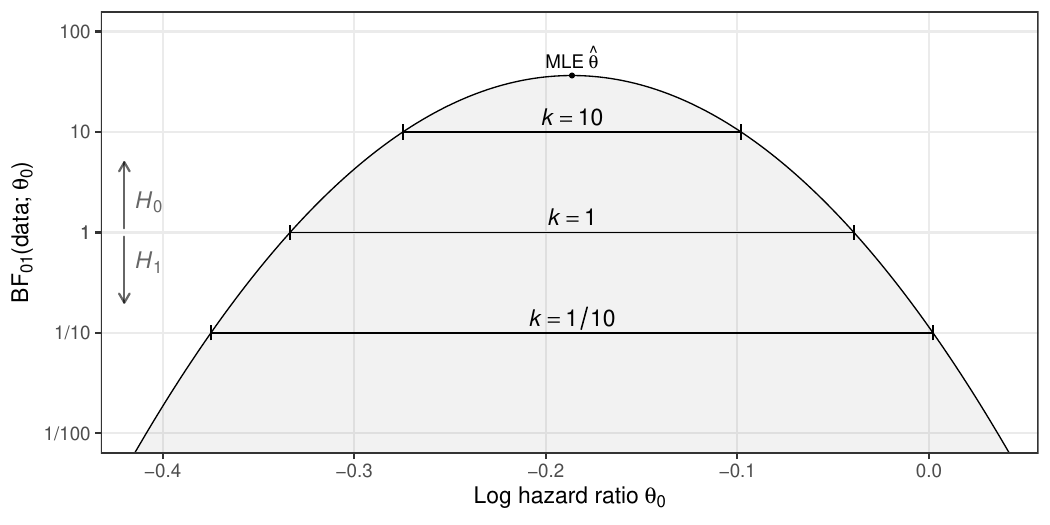} 
\end{knitrout}

\caption{The RECOVERY trial \citep{RECOVERY2021} found that dexamethasone
  treatment reduced mortality compared to usual care in hospitalized Covid-19
  patients (estimated log hazard ratio $\that = -0.19$ with
  standard error $\sigma = 0.05$ and 95\% confidence
  interval from $-0.29$ to
  $-0.07$). Assuming a normal likelihood
  $\that \given \theta \sim \Nor(\theta, \sigma^{2})$, the Bayes factor for
  contrasting $\h{0}\colon \theta = \theta_0$ to
  $\h{1}\colon \theta \neq \theta_0$ is shown as a function of the null value
  $\theta_0$. A unit-information normal distribution
  \mbox{$\theta \given \h{1} \sim \Nor(\mu_\theta = -0.22, \sigma_\theta^{2} = 4)$}
  centered around the clinically relevant log hazard ratio is used as prior for
  $\theta$ under $\h{1}$. Support intervals for different support levels $k$
  indicate the range of log hazard ratios supported by the data.}
    \label{fig:example}
\end{figure*}

Figure~\ref{fig:example} illustrates different support sets (in this case
intervals) for a log hazard ratio parameter $\theta$ quantifying the effect of
the drug dexamethasone on the mortality of hospitalized patients with Covid-19
enrolled in the RECOVERY trial \citep{RECOVERY2021}. Shown is also the Bayes
factor for testing $\h{0}\colon \theta = \theta_0$ versus
$\h{1}\colon \theta \neq \theta_0$ viewed as a function of the null value
$\theta_0$. A $k$ support set is obtained from ``cutting'' this function at
height $k$, and taking the parameter values with a Bayes factor value larger
than $k$ as part of the set. In practice, it is not clear which value of $k$
should be chosen. One possibility is to select $k$ based on conventional
classifications of Bayes factors or likelihood ratios. Table~\ref{tab:evidence}
lists three of them. For instance, using the classification from \citet[Appendix
B]{Jeffreys1961}, the $k=10$ support interval ranging from
$-0.27$ to $-0.1$ can be
interpreted to contain log hazard ratios that are \emph{strongly supported} by
the data, whereas the $k=1/10$ support interval ranging from
$-0.37$ to $0$ can be
interpreted to contain log hazard ratios that are \emph{at least not strongly
  contradicted} by the data.

\begin{table*}[!htb]
    \centering
    \caption{Classifications of evidence for $\h{0}$ provided by Bayes factors
      $\BF_{01} = k$. The cut-offs from Jeffreys are slightly adjusted from
      powers of $\surd 10$, as suggested by \citet{Held2018}. Royall and Fisher
      defined their classifications only for likelihood ratios, i.e., Bayes
      factors with simple hypotheses $\h{0}\colon\theta=\theta_{0}$ vs.
      $\h{1}\colon\theta=\theta_{1}$. While Royall placed no restrictions on
      $\theta_{1}$, Fisher used the maximum likelihood estimate
      $\theta_{1} = \that$. He named only the $k < 1/15$ category.}
    \label{tab:evidence}
    \resizebox{\columnwidth}{!}{%
    \begin{tabular}{c c c c c c}
      \toprule
      $k$ & \citet{Jeffreys1961} & $k$ & \citet{Royall1997} & $k$ & \citet{Fisher1956} \\
      \cmidrule(lr){1-2}
      \cmidrule(lr){3-4}
      \cmidrule(lr){5-6}
      > 100 & Decisive & > 64 & Quite strong indeed & 1/2 to 1 & Good \\
      30 to 100 & Very strong & 32 to 64 & Quite strong & 1/5 to 1/2 & Fair \\
      10 to 30 & Strong & 8 to 32 & Strong & 1/15 to 1/5 & Poor \\
      3 to 10 & Substantial & 4 to 8 & Weak & < 1/15 & Open to grave suspicion \\
      1 to 3 & Bare mention & & & & \\
      \bottomrule
    \end{tabular}%
    }
\end{table*}

The construction of support sets thus parallels the construction of frequentist
confidence sets: A $(1 - \alpha)100\%$ confidence set corresponds to the set of
parameter values which are not rejected by a null hypothesis significance test
at level $\alpha$. It can equally be displayed and obtained from a so-called
\emph{$p$-value function}, which is the $p$-value of the data viewed as a
function of the null value \citep{Fraser2019, Rafi2020}. Despite these
similarities, the interpretation of support and confidence sets is rather
different; support sets contain parameter values for which there is at least a
certain amount of statistical evidence, whereas confidence sets are defined
through the long-run frequency of including the unknown parameter $\theta$ with
probability equal to their confidence level. The parameter values in a
confidence sets are typically interpreted as being ``compatible'' with a
particular data set, but this is debatable as the confidence level is concerned
with the confidence set as a procedure over multiple replications.
% ``the infrequency with which, in particular circumstances, decisive evidence
% is obtained, should not be confused with the force, or cogency, of such
% evidence'' \citep[p. 93]{Fisher1956}.

Although support sets are conceptually simple and intuitive, they have not been
applied to many problems. It is also unclear how they relate to the more widely
used confidence sets. In this article we thus shed light on the connection
between support and confidence sets. Specifically, we provide methods for
calibrating approximate confidence sets to approximate support sets and vice
versa in the important case when the data consists of an estimate of a
univariate parameter $\theta$ with approximate normal likelihood
(Section~\ref{sec:SInormality}). To do so, we derive novel and easy-to-use
formulas for computing support intervals that only require summary statistics
typically reported in research articles, \eg{} point estimates, standard errors,
or confidence intervals. This scenario is highly relevant as a large part of
commonly used estimators satisfy the approximate normality assumption, and also
because one often does not have access to the raw data but only the summary
statistics. Computing a support interval requires the specification of a prior
distribution for $\theta$ under the alternative $\h{1}$, and we compare several
classes of distributions. We also show how bounding the evidence against the
null hypothesis for a certain class of prior distributions leads to the novel
concept of a \emph{minimum support set}. Our minimum support sets are directly
related to well-known bounds of Bayes factors \citep{Berger1987, Selke2001,
  Held2018}. In Section~\ref{sec:SIbounds}, we show how minimum support sets
provide confidence sets an evidential interpretation with respect to certain
classes of priors. We then illustrate how the sample size of a future study can
be determined based on support, which provides a novel alternative to the
conventional approaches based on either power or precision of an interval
estimator (Section~\ref{sec:design}). Finally, we show how the universal bound
for the type I error rate of Bayes factors can be used for bounding the coverage
of support sets, even under sequential analyses with optional stopping
(Section~\ref{sec:t1e}). As a running example, we use data from the RECOVERY
trial \citep{RECOVERY2021}, as already introduced in Figure~\ref{fig:example}.

\section{Support intervals under normality}
\label{sec:SInormality}
Denote by $\that$ an asymptotically normal estimator of an unknown univariate
parameter $\theta$, possibly the maximum likelihood estimator (MLE). Suppose its
squared standard error $\sigma^2$ is an estimate of the asymptotic variance of
$\that$, so that an approximate normal likelihood
$\that \given \theta \sim \Nor(\theta, \sigma^{2})$ is justifiable. For example,
$\that$ could be an estimated regression coefficient from a generalized linear
model and $\sigma$ its standard error. In many simple settings, the standard
error is of the form $\sigma = \lambda/\surd{n}$, where $\lambda^2$ is the
variance corresponding to one effective unit and $n$ is the effective sample
size, for example, the number of measurements or the number of events
\citep[Section 2.4]{Spiegelhalter2004}, see also \citet{Berger2013} for a
generalization of effective sample size to more complex settings with dependent
data. An approximate $(1 - \alpha)100\%$~confidence interval for $\theta$ is
given by
\begin{align}
    \label{eq:ci}
  \that \pm \sigma \times \Phi^{-1}(1 - \alpha/2)
\end{align}
with $\Phi^{-1}(\cdot)$ the quantile function of the standard normal
distribution. The confidence level $(1 - \alpha)100\%$ represents the long run
frequency with which the true parameter is included in the confidence interval
(assuming that the sampling model is correct). Note that the
interval~\eqref{eq:ci} also corresponds to the $(1 -\alpha)100\%$ posterior
credible interval based on an (improper) uniform prior for $\theta$,
corresponding to Jeffreys's transformation invariant prior \citep{Jeffreys1961,
  Ly2017} and thus also representing the default interval estimate for $\theta$
from a Bayesian estimation perspective. We will now contrast the confidence
interval~\eqref{eq:ci} to several types of support intervals.

\subsection{Normal prior under the alternative}
To construct a support interval for $\theta$ using the data summary $\that$ with
$\that \given \theta \sim \Nor(\theta, \sigma^{2})$, specification of a prior
for $\theta$ under the alternative $\h{1}$ is required. Specifying a normal
prior $\theta \given \h{1} \sim \Nor(\mu_\theta, \sigma_\theta^2)$ results in
the Bayes factor
% %% uncomment for short equation
% \begin{align}
%     \label{eq:bf01}
%   \BF_{01}(\that; \theta_0)
%   = \frac{\sqrt{1 + \sigma_\theta^2/\sigma^2}}{\exp\bigg[\dfrac{1}{2}\bigg\{\dfrac{(\that -
%   \that_0)^2}{\sigma^2} - \dfrac{(\that - \mu_\theta)^2}{\sigma^2 + \sigma_\theta^2}\bigg\} \bigg]}.
% \end{align}
%% uncomment for long equation
\begin{align}
    \label{eq:bf01}
  \BF_{01}(\that; \theta_0)
  = \sqrt{1 + \frac{\sigma_\theta^2}{\sigma^2}} ~ \exp\left[-\frac{1}{2}\left\{\frac{(\that -
  \theta_0)^2}{\sigma^2} - \frac{(\that - \mu_\theta)^2}{\sigma^2 + \sigma_\theta^2}\right\} \right].
\end{align}
Now, fixing the Bayes factor~\eqref{eq:bf01} to $k$ and solving for $\theta_0$ leads
to the $k$ support interval
\begin{align}
    \label{eq:si}
    \that \pm \sigma \times
    \sqrt{\log\left(1 + \frac{\sigma_\theta^2}{\sigma^2}\right) + \frac{(\that - \mu_\theta)^2}{\sigma^2 + \sigma_\theta^2} -
    2\log k}.
\end{align}

Similar to the confidence interval~\eqref{eq:ci}, the support
interval~\eqref{eq:si} is centered around the parameter estimate $\that$.
However, while the width of the confidence interval is only determined through
the confidence level $(1 - \alpha)100\%$ and standard error $\sigma$, the
width of the support interval also depends on the specified prior for $\theta$
under $\h{1}$. Moreover, for $k > 1$ it may happen that the support interval
is empty, as the term below the square root in~\eqref{eq:si} may become
negative for too large $k > 1$. This means that in order to find the desired
level of support $k > 1$, the data have to be sufficiently informative (relative
to the prior), \ie{} the squared standard error $\sigma^2$ has to be
sufficiently small relative to the prior variance $\sigma_\theta^2$.

In the following, we will discuss how different prior means $\mu_\theta$ and
variances $\sigma_\theta^2$ affect the resulting support intervals. When the
prior variance decreases ($\sigma_\theta^2 \downarrow 0$), the prior approaches
a point mass at $\mu_\theta$. The width of the support interval is then fully
determined by the difference between the parameter estimate $\that$ and the
prior mean $\mu_\theta$ divided by the standard error $\sigma$. A smaller
difference between $\that$ and $\mu_{\theta}$ leads to a tighter support
interval. In contrast, for priors that become increasingly diffuse
($\sigma_\theta^{2} \to \infty$), the $k \geq 1$ support interval~\eqref{eq:si}
extends to the entire real line, indicating that all values $\theta \in \R$
receive more support from the data than the diffuse alternative, regardless of
the data, i.e., the observed estimate $\that$, standard error $\sigma$, and the
location of the prior mean $\mu_\theta$. This particular behavior provides
another perspective on the well-known Jeffreys-Lindley paradox
\citep{Wagenmakers2021a}; the confidence interval from~\eqref{eq:ci} only spans
a finite range around the parameter estimate $\that$, so that the corresponding
null hypothesis significance tests would reject the parameter values outside,
whereas for the same values the Bayes factor would indicate evidence for the
null hypothesis. Finally, centering the prior around the parameter estimate
($\mu_\theta = \that$) and setting the prior variance equal to the variance of
one effective observation ($\sigma_\theta^2 = n \times \sigma^2$ with $n$ the
effective sample size), produces the support interval for Jeffreys's approximate
Bayes factor \citep{Wagenmakers2022} which is equal to the well-known
approximation of the Bayes factor based on the Bayesian information criterion
\citep{Raftery1999}. In this case, the standard error multiplier has a
particularly simple form ${\mbox{M} = \surd\{\log(1 + n) - 2 \log k\}}$, showing
that at least $n \geq k^2 - 1$ effective observations are required for the
respective support interval with $k \geq 1$ to be non-empty.

\subsection{Local normal prior under the alternative}
\label{sec:local}
The support interval based on the normal prior~\eqref{eq:si} depends on the
specification of a prior mean and prior variance. A different approach is to use
a so-called \emph{local prior}, that is, a unimodal and symmetric prior centered
around the null value $\theta_0$ \citep{Berger1987b}.
% , for instance, the unit-information prior
% $\theta \given \h{1} \sim \Nor(\mu_\theta = \theta_0, \sigma_\theta^2 = n \times \sigma^2)$ from
% \citet{Kass1995b}.
Choosing a local normal prior with variance $\sigma^2_\theta$ corresponds to
setting $\mu_\theta = \theta_0$ in~\eqref{eq:bf01}, which leads to the Bayes
factor
% %% uncomment for short equation
% \begin{align}
%     \label{eq:bf01local}
%   \BF_{01}(\that ; \theta_0)
%   = \frac{\sqrt{1 + \sigma_\theta^2/\sigma^2}}{
%   \exp\bigg\{\dfrac{1}{2} \, \dfrac{(\that - \theta_0)^2}{\sigma^2(1 + \sigma^2/\sigma_\theta^2)}
%   \bigg\}}.
% \end{align}
%% uncomment for long equation
\begin{align}
    \label{eq:bf01local}
  \BF_{01}(\that ; \theta_0)
  = \sqrt{1 + \frac{\sigma_\theta^2}{\sigma^2}}
  ~ \exp\left\{-\frac{1}{2} \, \frac{(\that - \theta_0)^2}{\sigma^2(1 + \sigma^2/\sigma_\theta^2)} \right\}.
\end{align}
The $k$ support interval based on the Bayes factor~\eqref{eq:bf01local} is then
given by
\begin{align}
    \label{eq:silocal}
    \that \pm \sigma \times
    \sqrt{\left\{\log\left(1 + \frac{\sigma_\theta^2}{\sigma^{2}}\right) - 2\log k \right\}
    \left(1 + \frac{\sigma^{2}}{\sigma_\theta^2}\right)}.
\end{align}

While the Bayes factor~\eqref{eq:bf01local} is a special case of the Bayes
factor~\eqref{eq:bf01}, the support interval~\eqref{eq:silocal} is not a special
case of the support interval~\eqref{eq:si}. This is because the prior for
$\theta$ under $\h{1}$ is different for each null value $\theta_0$, whereas it
is always the same under the two-parameter normal prior approach. To fully
specify the support interval~\eqref{eq:silocal}, the prior variance
$\sigma_\theta^{2}$ needs to be chosen. One standard choice is to set it equal
to the variance of a single observation ($\sigma_\theta^2 = n \times \sigma^2$),
known as unit-information prior \citep{Kass1995b}. This approach leads to the
$k$ support interval
\begin{align}
    \label{eq:silocalui}
    \that \pm \sigma \times
    \sqrt{\left\{\log\left(1 + n\right) - 2\log k \right\}
    \left(1 + 1/n\right)}.
\end{align}
For this type of support interval, the standard error multiplier
$\mbox{M} = \surd[\{\log(1 + n) - 2 \log k\}(1 + 1/n)]$ is wider than for the
Jeffreys's approximate Bayes factor by a factor of $\surd(1 + 1/n)$ but the
condition $n \geq k^2 - 1$ for the $k \geq 1$ support interval to be non-empty
is the same.

\subsection{Nonlocal normal moment prior under the alternative}
\label{sec:nonlocal}
Another attractive class of priors for $\theta$ under the alternative is given by so-called \emph{nonlocal priors}. These priors are characterized by having zero density at the null value $\theta_0$, thereby leading to a faster accumulation of evidence than local priors when the null hypothesis is actually true \citep{Johnson2010}. One popular type of nonlocal priors is given by \emph{normal moment priors} $\theta
\sim \mathrm{NM}(\theta_0,
\sigma_\theta)$, with symmetry point $\theta_0$ and spread $\sigma_\theta$ which have density ${f(\theta
  \given \theta_0, \sigma_\theta) = \Nor(\theta \, ; \, \theta_0,
  \sigma_\theta^2) \times (\theta -
  \theta_0)^{2}/\sigma_\theta^2}$ where $\Nor(\cdot \, ; \, \theta_0,
\sigma_\theta^2)$ denotes the density function of a normal distribution with mean $\theta_0$ and variance $\sigma_\theta^2$. The Bayes factor employing a prior $\theta
\given \h{1} \sim \mathrm{NM}(\theta_0, \sigma_\theta)$ is then given by
% %% uncomment for short equation
% \begin{align*}
%   \BF_{01}(\that ; \theta_0)
%   =&
%      \left(1 + \frac{\sigma_\theta^2}{\sigma^2}\right)^{3/2} \,
%      \exp\left\{-\frac{1}{2} \, \frac{(\that - \theta_0)^2}{\sigma^2(1 + \sigma^2/\sigma_\theta^2)}
%      \right\} \\
%    & \times \left(1 + \frac{(\that - \theta_0)^2}{\sigma^2(1 + \sigma^2/\sigma_\theta^2)}\right)^{-1}
%      \nonumber
% \end{align*}
%% uncomment for long equation
\begin{align*}
  \BF_{01}(\that ; \theta_0)
  = \left(1 + \frac{\sigma_\theta^2}{\sigma^2}\right)^{3/2} \,
  \exp\left\{-\frac{1}{2} \, \frac{(\that - \theta_0)^2}{\sigma^2(1 + \sigma^2/\sigma_\theta^2)} \right\}
  \left\{1 + \frac{(\that - \theta_0)^2}{\sigma^2(1 + \sigma^2/\sigma_\theta^2)}\right\}^{-1}
\end{align*}
from which the corresponding $k$ support interval can be derived to be
% %% uncomment for short equation
% \begin{align}
%   \label{eq:sinonlocal}
%   \that \pm \sigma \times
%   \sqrt{\dfrac{2 \lw{0}\left\{\dfrac{(1 + \sigma_\theta^2/\sigma^2)^{3/2}
%   \sqrt{e}}{2 k}\right\} - 1}{
%   \left(1 + \sigma^2/\sigma^2_\theta\right)^{-1}}}
% \end{align}
%% uncomment for long equation
\begin{align}
  \label{eq:sinonlocal}
  \that \pm \sigma \times
  \sqrt{\left[2 \lw{0}\left\{\frac{(1 + \sigma_\theta^2/\sigma^2)^{3/2} \sqrt{e}}{2 k}\right\} - 1\right]
  \left(1 + \frac{\sigma^2}{\sigma_\theta^2}\right)}
\end{align}
with $\lw{0}(\cdot)$ denoting the principal branch of the Lambert W function.
The Lambert W function is the (complex) multivalued function $\text{W}(\cdot)$
satisfying $\text{W}(x)\, \exp\{\text{W}(x)\} = x$. For real $x$, it is defined
for $x \in [-1/e,\infty)$. For $x \geq
0$ the function has a unique value, whereas in the interval $x \in (-1/e,
0)$, the function has two branches: $\lw{0}(x) > -1$ for all $x \in (-1/e,
0)$ termed the principal branch, and $\lw{-1}(x) < -1$ for all $x \in (-1/e,
0)$, see \citet{Corless1996} for more details. It is possible that the support interval~\eqref{eq:sinonlocal} is empty, as for the other two types of support intervals. This happens when the Lambert W term is smaller than one half so that the square root is undefined. Since $\lw{0}(0.82) \approx 1/2$, this situations occurs when \mbox{$(1 +
\sigma_\theta^2/\sigma^2)^{3/2} < 0.82 \times
2k\sqrt{e}$}, meaning that the standard error $\sigma$ has to be sufficiently small relative to the prior spread parameter $\sigma_\theta$ and the support level $k$, so that the interval is non-empty.

\begin{figure}[!htb]
\begin{knitrout}
\definecolor{shadecolor}{rgb}{0.969, 0.969, 0.969}\color{fgcolor}
\includegraphics[width=\maxwidth]{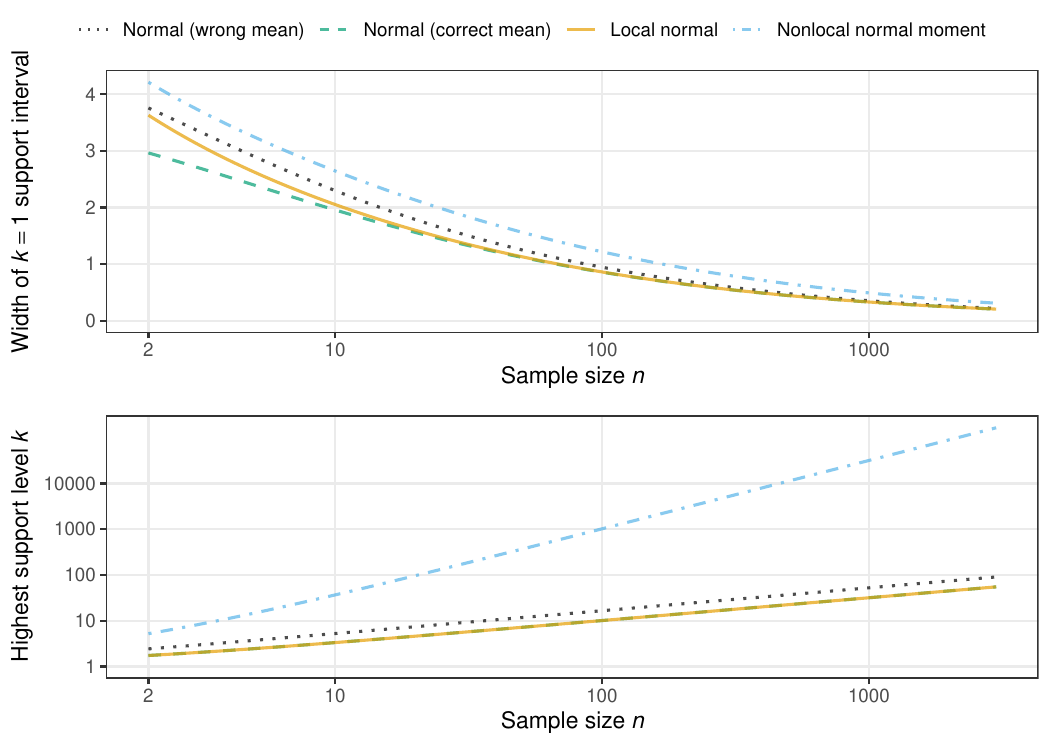} 
\end{knitrout}

\caption{Comparison of prior distributions for the parameter $\theta$ under the
  alternative $\h{1}$ in terms of the resulting support interval width and the
  highest level for which it is non-empty. A data model
  \mbox{$\that \given \theta \sim \Nor(\theta, \lambda^{2}/n = 4/n)$}
  is assumed in all cases. The prior scale/spread parameter is set to
  $\sigma_{\theta} = 2$. The normal prior (correct mean) has
  a mean equal to the parameter estimate $\that$, while the normal prior (wrong
  mean) has a mean one standard deviation $\lambda = 2$ away
  from $\that$.}
\label{fig:priorComparison}
\end{figure}

\subsection{Comparison of priors}
To better understand the advantages and disadvantages of the previously
discussed priors, the resulting support intervals
% priors based on normal priors~\eqref{eq:si}, local normal
% priors~\eqref{eq:silocal}, and nonlocal normal priors~\eqref{eq:sinonlocal}
can be compared in terms of their width as a function of the sample size $n$
(Figure~\ref{fig:priorComparison} top). For small sample sizes, the normal prior
with mean equal to the observed parameter estimate produces the narrowest
$k = 1$ support intervals, followed by the local normal prior, the normal prior
with mean one standard deviation away from the observed estimate, and lastly the
nonlocal normal moment prior. Thus, a well-chosen normal prior can increase the
precision of support inference, whereas a poorly chosen normal prior can
decrease precision. However, the differences in width between the priors mostly
disappear with increasing sample size. In the realistic range between ten and a
few hundred samples, the local normal prior seems to be a reasonable default
choice, as it leads to support intervals almost as narrow as the normal (correct
mean) prior, without the need to specify a mean.

Another aspect in which the priors can be compared is the highest support level
$k$ for which the resulting support intervals are non-empty
(Figure~\ref{fig:priorComparison} bottom). We see that for the same sample size
$n$, the highest support levels from the normal and local normal priors are
similar and show the same growth rates. In contrast, the highest support level
from the nonlocal moment prior is higher and grows much faster. This is expected
because nonlocal priors are designed to produce Bayes factors with faster
accumulation of evidence for the null hypothesis. Thus, although nonlocal moment
priors result in wider support intervals than the other priors, for small sample
sizes they may be the only type of prior that can produce a support interval at,
say, Jeffreys's strong evidence level $k = 10$.

\section{Support intervals based on Bayes factor bounds}
\label{sec:SIbounds}
In some situations it is clear which prior for $\theta$ should be chosen under
the alternative $\h{1}$, \eg{} when a parameter estimate from a previous data
set is available. In other situations it is less clear and different priors may
produce drastically different results. To provide a more objective assessment of
evidence in the latter situation, several authors have proposed to instead
specify only a class of prior distributions and then select the one prior among
them that leads to the Bayes factor providing the strongest possible evidence
against the null hypothesis $\h{0}$ \citep{Edwards1963, Berger1987, Selke2001,
  Held2018}. Here we refer to these Bayes factor bounds as \emph{minimum Bayes
  factors} for the null $\h{0}$ over the alternative $\h{1}$, as we are
interested in the support for null values $\theta_0$.

We will now show how minimum Bayes factors can be used for obtaining so-called
\emph{minimum support sets}. Specifically, a $k$ minimum support set is given by
\begin{align}
  \mbox{minSI}_k =
  \left\{\theta_0 : \minBF_{01}(x ; \theta_0) \geq k \right\},
\end{align}
where $\minBF_{01}(x ; \theta_0)$ is the smallest possible Bayes factor for
testing $\h{0}\colon \theta = \theta_0$ versus
${\h{1}\colon \theta \neq \theta_0}$ that can be obtained from a class of prior
distributions for $\theta$ under the alternative $\h{1}$. That is, given the
data, for each $\theta_0$ the prior for $\theta$ under $\h{1}$ is cherry-picked
from a class of priors to obtain the lowest evidence for
$\h{0}\colon \theta = \theta_0$ possible. Minimum support intervals thus provide
a Bayes/non-Bayes compromise \citep{Good1992} as they do not require
specification of a specific prior distribution but still allow for an evidential
interpretation of the resulting interval.

One property of minimum Bayes factors is that they can only be used to asses the
maximum evidence \emph{against} the null hypothesis but not for it. Minimum
support sets inherit this property, meaning that they can only be obtained for
support levels $k \leq 1$. For instance a $k = 1/3$ minimum support set includes
the parameter values under which the observed data are \emph{at most} $3$ times
less likely compared to under all priors from the specified class of
alternative. Being unable to obtain support intervals with $k > 1$ is the price
that needs to be paid for having to only specify a class of prior distributions
but not a specific prior itself. We will now discuss minimum support intervals
from several important classes of distributions.

\subsection{Class of all distributions under the alternative}

Among the class of all possible priors under $\h{1}$, the prior which is most
favorable towards the alternative is a point mass at the observed effect
estimate $\h{1}\colon \theta = \that$ \citep{Edwards1963}. The resulting minimum
Bayes factor is given by
\begin{align}
    \label{eq:minBFsimple}
  \minBF_{01}(\that ; \theta_0) =
  \exp\left\{-\frac{1}{2}  \frac{(\that - \theta_0)^{2}}{\sigma^{2}}\right\},
\end{align}
for which twice the negative log equals the standard likelihood ratio test
statistic when $\that$ is the MLE. Inverting~\eqref{eq:minBFsimple} for
$\theta_0$ leads to the $k$ minimum support interval
\begin{align}
    \label{eq:sisimple}
  \that \pm \sigma \times \sqrt{-2\log k}.
\end{align}
Interestingly, defining a support interval relative to the likelihood of the
data under the MLE has already been suggested by \citet{Fisher1956}.
Table~\ref{tab:evidence} shows Fisher's classification of evidence for this type
of interval. Also Royall made use of the minimum support
interval~\eqref{eq:sisimple}, usually with support levels $k=1/8$ and $k=1/32$.
He noted: ``The $1/8$ and $1/32$ likelihood intervals are not confidence
intervals, in general, but they truly represent what confidence intervals are
often mistaken to represent, namely parameter values that the sample does not
represent evidence against, that is, values that are `consistent with the
observations'. We can speak in this way, asserting that there is not strong
evidence against a point inside the interval, without reference to an
alternative value, because the statement is true for all alternatives. Every
point inside the $1/8$ interval is consistent with the observations in the
strong sense that there is no other possible value of the parameter that is
better supported by a factor as large as $8$'' \citep[p. 101]{Royall1997}. While
we agree that the support interval~\eqref{eq:sisimple} is a useful bound, it is
important to note that from a Bayesian perspective it represents the most
blatantly biased assessment of support in the sense that assigning a point prior
at the observed parameter estimate hardly reflects prior knowledge about
$\theta$ but can rather be considered cheating \citep{Berger1987}. This is
reflected by the fact that for a given estimate (i.e., data set) and fixed
support level $k$, the interval represents the narrowest support interval among
all possible support intervals. When minimizing over the class of all
two-parameter normal priors, i.e., the Bayes factor~{\eqref{eq:bf01}}, we also
obtain the same minimum Bayes factor~{\eqref{eq:minBFsimple}} and consequently
the same minimum support interval~\eqref{eq:sisimple}.

\subsection{Class of local normal alternatives}

When the class of priors for $\theta$ under the alternative $\h{1}$ is given by
normal distributions centered around the null value $\theta_0$, choosing its
variance to be $\sigma_\theta^2 = \max\{(\that - \theta_0)^2 - \sigma^2, 0\}$
maximizes the marginal likelihood of the data under $\h{1}$. Plugging this
variance in the Bayes factor~\eqref{eq:bf01local} leads to the minimum Bayes
factor over the class of local normal priors
%% uncomment for long equation
\begin{align}
  \minBF_{01}(\that ; \theta_0) =
  \begin{cases}
    \dfrac{|\that - \theta_0|}{\sigma} \exp\left\{-\dfrac{(\that -
    \theta_0)^2}{2\sigma^{2}}\right\}  \sqrt{e}
    & ~ \text{if} ~ \dfrac{|\that - \theta_0|}{\sigma} > 1 \\
    1 & ~ \text{else}
  \end{cases}
  \label{eq:minBFnorm}
  \end{align}
% %% uncomment for short equation
% \begin{align}
%   \minBF_{01}(\that ; \theta_0) =
%   \begin{cases}
%     |z_{\scriptscriptstyle \theta_0}| \, \exp\left\{
%     -\dfrac{z_{\scriptscriptstyle \theta_0}^2 - 1}{2}\right\}
%     & ~ \text{if} ~ z_{\scriptscriptstyle \theta_0} > 1 \\
%     1 & ~ \text{else}
%   \end{cases}
%   \label{eq:minBFnorm}
% \end{align}
% with $z_{\scriptscriptstyle \theta_0} = (\that - \theta_0)/\sigma$,
as first shown by \citet{Edwards1963}. Equating~\eqref{eq:minBFnorm} to $k$ and
solving for $\theta_0$ leads then to the $k$ minimum support interval
\begin{align}
  \that \pm \sigma \times \sqrt{-\lw{-1}(-k^2/e)},
\end{align}
with $\lw{-1}(\cdot)$ the branch of the Lambert W function that satisfies
$\mbox{W}(y) < -1$ for $y \in (-e^{-1}, 0)$. For $k = 1$, the standard error
multiplier becomes $\mbox{M} = \sqrt{-\lw{-1}(-1/e)} = 1$. Hence, the data
provide support for all parameter values within one standard error around the
observed parameter estimate $\that$ when the class of priors for the parameter
is given by local normal alternatives.

\subsection{Class of \textit{p}-based alternatives}

\citet{Vovk1993} and \citet{Selke2001} proposed a minimum Bayes factor where the
data are summarized through a $p$-value. The idea is that under the null
hypothesis $\h{0}\colon \theta = \theta_0$, a $p$-value should be uniformly
distributed, whereas under the alternative it should have a monotonically
decreasing density characterized by the class of Beta($\xi, 1$) distributions
(with $\xi \leq 1$). Choosing $\xi$ such that the marginal likelihood of the
data under $\h{1}$ is maximized, leads to well-known ``$-ep \log p$'' minimum
Bayes factor
\begin{align}
  \minBF_{01}(p ; \theta_0) =
  \begin{cases}
    - e  p  \log p
    & ~ \text{if} ~ p \leq e^{-1} \\
    1 & ~ \text{else}
  \end{cases}
  \label{eq:minBFeplog}
\end{align}
with $p = 2\{1 - \Phi(|\that - \theta_0|/\sigma)\}$.
Equating~\eqref{eq:minBFeplog} to $k$ and solving for $\theta_0$, leads to the
$k$ minimum support interval
\begin{align}
  \that \pm \sigma \times \Phi^{-1}\left[1 -
  \frac{\exp\left\{\lw{-1}(-k/e)\right\}}{2}\right].
\end{align}
For $k = 1$, the standard error multiplier is given by
$\mbox{M} = \Phi^{-1}[1 - \exp\{\lw{-1}(-1/e)\}/2] = {\Phi^{-1}[1 - 1/(2e)]} \approx 0.90$,
so the $k = 1$ minimum support interval is just slightly tighter than the one
based on local normal alternatives.

%% place at appropriate place in preprint/journal

\subsection{Mapping between confidence and minimum support levels}
\begin{figure*}[!tb]
\begin{knitrout}
\definecolor{shadecolor}{rgb}{0.969, 0.969, 0.969}\color{fgcolor}
\includegraphics[width=\maxwidth]{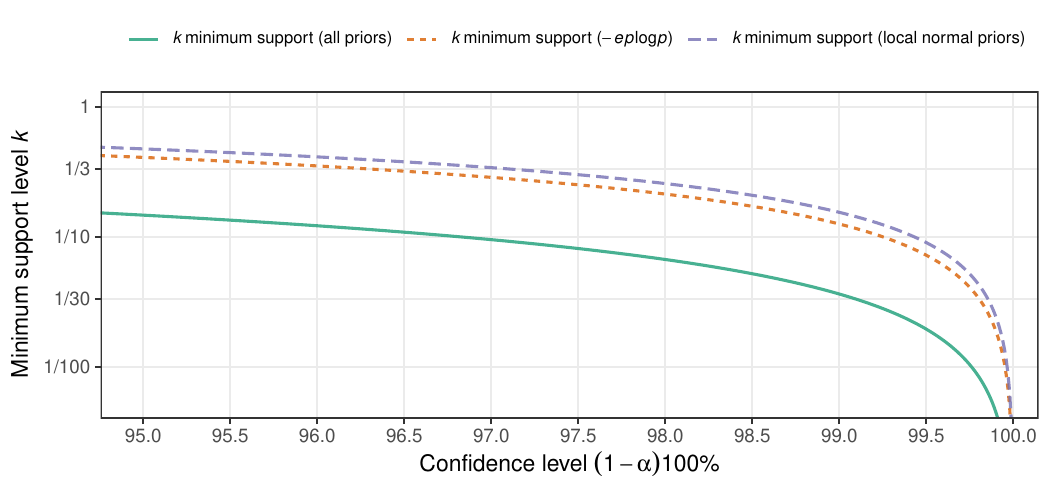} 
\end{knitrout}
\caption{Mapping between confidence level $(1 - \alpha)100\%$ and minimum
  support level $k$ for different types of minimum support intervals.}
    \label{fig:mappingCIk}
  \end{figure*}

For all types of minimum support intervals discussed so far, there is a
one-to-one mapping between their minimum support level $k$ and the confidence
level $(1 - \alpha) 100\%$ of the approximate confidence interval~\eqref{eq:ci},
see Figure~\ref{fig:mappingCIk}. The conventional default level of 95\%
corresponds to a $k = 1/6.8$ support level for the class of
all priors under the alternative, a $k = 1/2.5$ support level
for the $-ep\log p$, and a $k = 1/2.1$ support level for the
local normal prior calibration. Conversely, the $k = 1/10$ minimum support
interval corresponds to the $96.81\%$ confidence
interval for the class of all priors, the $99.25$\%
confidence interval for $-ep\log p$, and the $99.43$\%
confidence intervals for the local normal prior calibration. Similar to the
mappings between Bayes factor bounds and $p$-values \citep{Held2018}, the
mappings displayed in Figure~\ref{fig:mappingCIk} provide confidence intervals
an evidential interpretation. Specifically, they enhance their long-term
frequency interpretation with an interpretation that directly relates to the
minimum support that the observed data provide for the parameter values in the
interval.

\section{Example RECOVERY trial}
We now compute the above (minimum) support intervals for the data from the
RECOVERY trial \citep{RECOVERY2021}. With the standard error $\sigma$ known, the
minimum support intervals are fully specified and can be readily computed. For
the normal, local normal, and the nonlocal normal moment prior we choose their
parameters as follows. The trial steering committee determined the sample size
of the trial based on an assumed clinically relevant log hazard ratio of
$\log 0.8 = -0.22$. This effect size can be used to
inform the normal prior under the alternative $\h{1}$, \ie{} we specify the mean
$\mu_\theta = -0.22$ along with the unit-information variance
$\sigma_\theta^2 = 4$ for a log hazard ratio \citep[Section
2.4.2]{Spiegelhalter2004}. Likewise, we use the unit-information variance
$\sigma_\theta^2 = 4$ as the variance of the local normal
prior. The spread parameter of the nonlocal moment prior $\sigma_{\theta}$ is
elicited with a similar approach as in \citet{Pramanik2022}; The value
$\sigma_\theta = 0.28$ is selected so that
90\% probability mass is assigned to log hazard
ratios between $\theta_0 - \log2$ and $\theta_0 + \log 2$, representing effect
sizes that at most half or double the mortality hazards relative to the null
value $\theta_0$.

\begin{figure*}[!htb]
\begin{knitrout}
\definecolor{shadecolor}{rgb}{0.969, 0.969, 0.969}\color{fgcolor}
\includegraphics[width=\maxwidth]{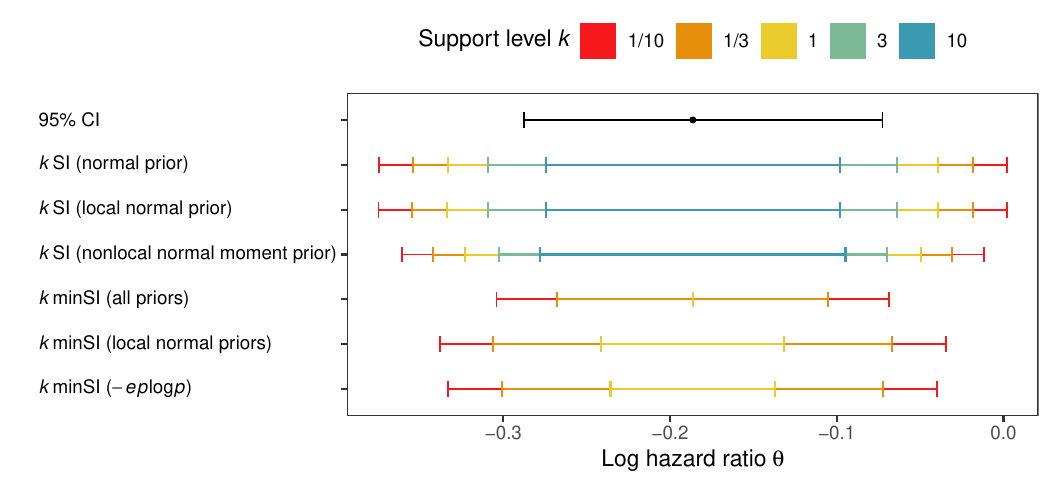} 
\end{knitrout}

\caption{Different support intervals for the data from the RECOVERY trial.
  The normal prior is centered around % the clinically relevant proportional
  % mortality reduction
  $\mu_{\theta} = % \log 0.8 =
  -0.22$
  % (as deemed by the trial steering committee)
  and has unit variance $\sigma_\theta^{2} = 4$. The local
  normal prior also has unit variance $\sigma_\theta^{2} = 4$.
  The spread parameter of the nonlocal normal moment prior is
  $\sigma_\theta = 0.28$. % so that
  % round(100*pnonlocal, 2)\% of the probability is assigned to log hazard
  % ratios in $[\theta_0 \pm \log2]$, representing effect sizes that at most half
  % or double the mortality hazards relative to the null value $\theta_0$.
}
    \label{fig:comparison}
\end{figure*}

Figure~\ref{fig:comparison} shows the corresponding $k$ support intervals for
different values of $k$. The support intervals based on normal (second row) and
local normal prior (third row) mostly coincide for all considered support levels
$k$.
% This is because the mean $m = -0.22$ of the prior under $\h{1}$ is close to the
% observed effect estimate $\that = -0.19$, so that
The $k = 10$ support intervals (blue) from both types indicate that log hazard
ratios between $-0.27$ and
$-0.1$ receive strong support from the data
compared to alternative parameter values. In contrast, the $k = 10$ support
interval (blue) based on the nonlocal normal moment prior (fourth row) is
slightly wider, indicating that values between
$-0.28$ and
$-0.09$ are strongly supported by the
data. For smaller support levels ($k < 10$) this trend reverses and the normal
and local normal prior support intervals are wider than the one based on the
nonlocal normal prior. Finally, each parameter value not included in a $k$
support interval corresponds to a point-null hypothesis for which the respective
Bayes factor is smaller than $k$, similar to the relationship between
confidence intervals and $p$-values. For instance, one can immediately see that
the Bayes factor based on nonlocal moment priors indicates strong evidence
($\BF_{01} < 1/10$) against $\h{0}\colon\theta = 0$ as the value is not included
in the interval, whereas this is not the case for the Bayes factors based on
normal and local normal priors.

The three bottom rows in Figure~\ref{fig:comparison} show different types of $k$
minimum support intervals computed for the data from the RECOVERY trial. Since
minimum support intervals are only non-empty for $k \leq 1$, only such support
levels are shown. The (yellow) $k=1$ minimum support interval for the class of
all priors (fifth row) is just a point at the observed effect estimate
$\that = -0.19$. In contrast, the (yellow) $k = 1$ minimum
support intervals based on local normal priors (sixth row) and the $-ep \log p$
calibration (last row) span about one standard error around the effect estimate.
Also for $k = 1/3$ (orange) and $k = 1/10$ (red), the minimum support interval
based on the class of all priors is much narrower than the ones based on
local normal and $-ep\log p$, yet all of them are narrower than the ordinary
support intervals. This illustrates that minimum support intervals provide an
overly pessimistic assessment of support for parameter values, in the same way
that Bayes factor bounds provide an overly pessimistic quantification of
evidence for the null hypothesis.

\section{Design of new studies based on support}
\label{sec:design}
The sample size of a future study is typically derived to achieve (i) a targeted
power of a hypothesis test, or (ii) a targeted precision of a future
confidence/credible interval. Here, we provide an alternative where the sample
size of a future study is determined to achieve a desired level of support.

Assume we wish to conduct a study and analyze the resulting parameter estimate
$\that$ using the support interval based on a normal prior~\eqref{eq:si}.
Further assume that we either specify a reasonable prior from existing knowledge
or use the prior for Jeffreys's approximate Bayes factor. The goal is now to
determine the sample size $n$ such that we can identify the parameter values
which are strongly supported by the future data, for instance, with a support
level $k = 10$ representing ``strong'' support in the classification from
\citet{Jeffreys1961}. In order for the $k > 1$ support interval~\eqref{eq:si} to
be non-empty, the standard error $\sigma$ of the parameter estimate $\that$
needs to be sufficiently small so that the term in the square root becomes
non-negative, \ie{} it must hold that
\begin{align}
  \label{eq:existence}
  \log\left(1 + \frac{\sigma_\theta^2}{\sigma^2}\right) + \frac{(\that - \mu_{\theta})^{2}}{\sigma^2 + \sigma_\theta^2} \geq 2 \log k.
\end{align}
The sample size $n$ can now be determined such that the standard error $\sigma$
is small enough for~\eqref{eq:existence} to hold. The resulting sample size then
guarantees that parameter values with the desired level of support will be
identified. In general, this needs to be done numerically, but for the
Jeffreys's approximate Bayes factor prior ($\mu_{\theta} = \that$ and
$\sigma_\theta^2 = n \sigma^2$), the simple expression $n \geq k^2 - 1$
mentioned earlier exists. For instance, if we want a $k = 10$ support interval
to be non-empty, we must take at least $10^2 - 1 = 99$ samples.

While the previously described approach guarantees that a $k > 1$ support
interval is non-empty and includes at least one parameter value $\theta$, one
may want to guarantee that the resulting $k$ support interval will span a
desired length
\begin{align}
    \label{eq:width}
    \ell = 2 \sigma \times \mbox{M}_k,
\end{align}
with $\mbox{M}_k$ the standard error multiplier of a $k$ support interval. In
general, numerical methods are required for computing the $n$ such
that~\eqref{eq:width} is satisfied, yet again for the support interval based on
Jeffrey's approximate Bayes factor
%($\mbox{M}_k = \{\log(1 + n) - 2\log k\}$)
there are explicit solutions available
\begin{align}
    \label{eq:nw}
    n = k^2 \, \exp\left\{-\lw{}\left(-\frac{k^2 \ell^2}{4 \lambda^2}\right)\right\}
\end{align}
with $\lambda^2$ the variance of one (effective) observation and assuming
$\log(1 + n)/\log(n) \approx 1$. From~\eqref{eq:nw} two things are apparent: (i)
the argument to $\mbox{W}(\cdot)$ has to be larger than $-1/e$ for the function
value to be defined, meaning that the possible width is limited by
$\ell \leq (4\lambda^2)/k^2$, (ii) since the argument to $\mbox{W}(\cdot)$ is
negative, there are always two solutions given by the two real branches of the
Lambert $\mbox{W}$ function, if any exist at all. For instance, for a standard
error of $\sigma = \lambda/\surd{n}$ with
$\lambda = 2$, a support level
$k = 10$, and a desired width $\ell = 0.2$,
equation~\eqref{eq:nw} leads to the sample sizes $n_1 = 143$
and $n_2 = 862$ (when rounded to the next larger integer).
Both lead to the $k = 10$ support interval spanning the desired
width $\ell = 0.2$, yet for the study employing the larger
sample size $n_2$ other support intervals with higher support levels $k$ can be
computed compared to a study employing the smaller sample size $n_1$.

\section{Error control via the universal bound}
\label{sec:t1e}
The universal bound \citep[Section 1.4]{Royall1997} ensures that for $k < 1$ and
when the null hypothesis $\h{0} \colon \theta = \theta_0$ is true, the
probability for finding evidence at most of level $k$ for $\h{0}$ cannot be
larger than $k$, that is
\begin{align}
    \label{eq:ubound}
  \P\left\{\BF_{01}(x ; \theta_0) \leq k \given \h{0}\right\} \leq k
\end{align}
for any prior of $\theta$ under the alternative $\h{1}$. Remarkably, the
universal bound is also valid under sequential analyses with optional stopping
as soon as a Bayes factor smaller than $k$ is obtained (\citet{Robbins1970,
  Pace2020}). In contrast, frequentist tests and confidence sets typically have
to be adjusted for sequential analyses to guarantee appropriate error rates, and
the theory and applicability can become quite involved.

\citet{Lindon2020} proved that $k$ support sets with $k < 1$ are also valid
$(1 - k)100\%$ confidence sets. Their proof and the related ``safe and anytime
valid inference'' theory \citep[see \eg{}][]{Grundwald2019} is based on
relatively technical results from martingale theory. We now briefly show how the
universal bound can also be used to derive error rate guarantees for support
intervals. Assume there is a true parameter $\theta = \theta_*$. For any
(data-independent) prior for $\theta$ under the alternative hypothesis $\h{1}$,
the coverage of the corresponding $k$ support set $\mbox{SI}_k$ with $k < 1$ is
bounded by
\begin{align}
    \P\left(\mbox{SI}_k \ni \theta_* \given \theta = \theta_*\right)
    &= \P\left\{\BF_{01}(x ; \theta_*) \geq k \given \theta = \theta_*\right\}  \nonumber \\
    &= 1 -  \P\left\{\BF_{01}(x ; \theta_*) < k \given \theta = \theta_*\right\} \nonumber \\
    &\geq 1 - k
    \label{eq:covbound}
\end{align}
where the first equality follows from the definition of a $k$ support
set~\eqref{eq:ss}, whereas the inequality follows from the universal
bound~\eqref{eq:ubound}. This shows that a $k$ support set with $k < 1$ is also
a valid $(1 - k)100\%$ confidence set, even under sequential analyses with
optional stopping, so that computing support intervals based on accumulating
data leads to a $(1 - k)100\%$ confidence sequence \citep{Lai1976, Howard2021}.
 Of course, the coverage bound rests on the assumption that
the data model is correctly specified and a misspecified data model will result
in incorrect coverage. Furthermore, the bound is based on simple null
hypotheses, but it can also be shown to hold for composite null hypotheses when
special types of priors are assigned to the nuisance parameters
\citep{Hendriksen2021}.

For the case of a univariate parameter $\theta$ as considered earlier,
construction of $(1 - k)100\%$ approximate confidence interval via the
normal prior support interval from~\eqref{eq:si} corresponds to the proposal by
\citet{Pace2020}. These authors studied this particular case in detail and gave
also frequentist motivations for the prior distributions interpreting them as
weighting functions. Moreover, they found that the method is also applicable to
parameter estimates from marginal, conditional, and profile likelihoods, and
that the coverage of the intervals is controlled even under slight model
misspecifications. We refer to \citet{Pace2020} for further details.

A $k < 1$ support interval will usually be wider than a standard $(1 - k)100\%$
confidence interval. On the other hand, a $k < 1$ support interval has at least
$(1 - k)100\%$ coverage, even under optional stopping (at least for point null
hypotheses as is the case here), which is not satisfied by a standard
$(1 - k)100\%$ confidence interval. Due to their property of valid coverage based
on arbitrary number of looks at the data, $k < 1$ support interval will
also typically be wider than $(1 - k)100\%$ confidence intervals adjusted via group
sequential or adaptive trial methodology which are more fine-tuned to specific
interim analysis strategies \citep{Wassmer2016}. These strategies are, however,
typically more restrictive and computationally involved compared to the flexible
and easily computable $k < 1$ support intervals which we present here.

It must be noted that the coverage bound~\eqref{eq:covbound} only holds for
support intervals but not for minimum support intervals. This is because the
minimum support intervals are derived based on priors that depend on the data,
which violates the assumption of the universal bound. Minimum support intervals
are thus only useful for giving confidence intervals an evidential
interpretation, but a $k$ minimum support interval with $k < 1$, itself does not
provide $(1 - k)100\%$ coverage under optional stopping.

\section{Discussion}
Misinterpretations and misconceptions of confidence intervals are common
\citep{Hoekstra2014, Greenland2016}. We showed how confidence intervals can be
reinterpreted as minimum support intervals which have an intuitive
interpretation in terms of the minimum evidence that the data provide for the
included parameter values. We also obtained easy-to-use formulas for different
types of support intervals for an unknown parameter based on an estimate and
standard error thereof. Table~\ref{tab:summary} summarizes our results, their
limitation being the reliance on the normality assumption which
may be inadequate for small sample sizes. More appropriate support intervals
can be obtained from considering the exact likelihood of the data instead of a
normal approximation, however, typically the support interval will not be available in
closed-form anymore and require the raw data rather than only the point estimate
and standard error.

%% place at appropriate place in preprint/journal
\begin{table*}[!htb]
    \centering
    \caption{Summary of confidence intervals (CI), support intervals (SI), and
      minimum support intervals (minSI) for an unknown parameter $\theta$ based
      on a parameter estimate $\that$ with standard error $\sigma$. All
      intervals are of the form $\that \pm \sigma \times \mbox{M}$. To transform
      an interval from type A to type B, first subtract $\that$ from the
      boundaries of the interval, multiply by the ratio of the standard error
      multipliers $\mbox{M}_{\text{B}}/\mbox{M}_{\text{A}}$, and add again
      $\that$ to the boundaries of the interval. The standard error multipliers
      $\mbox{M}$ depend on either the confidence level $(1 - \alpha)$ or the
      support level $k$. For the support intervals, the standard error
      multipliers M additionally depend on the parameters of the prior for
      $\theta$ under the alternative hypothesis: mean $\mu_\theta$ and variance
      $\sigma_\theta^{2}$ for the normal prior, variance $\sigma_\theta^{2}$ for
      the local normal prior, and spread $\sigma_\theta$ for the nonlocal normal
      moment prior. The quantile function of the standard normal distribution is
      denoted by $\Phi^{-1}(\cdot)$, $\lw{0}(\cdot)$ denotes the principal
      branch of the Lambert W function, and $\lw{-1}(\cdot)$ denotes the branch
      that satisfies $\mbox{W}(y) < 1$ for $y \in (-1/e, 0)$. (Minimum) support
      intervals are only non-empty for support levels $k$ for which the standard
      error multiplier is real-valued, \ie{} the term in the square root must be
      non-negative and/or the argument for $\lw{-1}(\cdot)$ must be in
      $[-1/e, 0)$. All interval types can be computed with the R package
      \texttt{ciCalibrate} (Appendix~\ref{app:Rpkg}).}
    \label{tab:summary}
    \begin{tabular}{lll}
    \toprule
    Interval type & %(Implied) prior for $\theta$ &
    Standard error multiplier M \\
    \midrule
    $(1 - \alpha)100\%$ CI %& $\theta \sim \Nor(\mu_\theta, \infty)$
    & $\Phi^{-1}(1 - \alpha/2)$ \vspace{0.2em}\\
    % $(1 - \alpha)100\%$ Credible & \theta \sim \Nor(\mu_\theta, \sigma_\theta^2) & \\
    $k$ $\mbox{SI}$ (normal prior) & %$\theta \sim \Nor(\mu_\theta, \sigma_\theta^2), ~ m \neq \theta_0$ &
    $\surd \{\log(1 + \sigma_\theta^2/\sigma^2) + (\that - \mu_\theta)^2/(\sigma^2 + \sigma_\theta^2) -
    2\log k\}$ \vspace{0.2em}\\
    $k$ $\mbox{SI}$ (local normal prior) & %$\theta \sim \Nor(\theta_0, \sigma_\theta^2)$ &
    $\surd [\{\log(1 + \sigma_\theta^2/\sigma^{2}) - 2\log k \}
    (1 + \sigma^{2}/\sigma_\theta^2)]$ \vspace{0.2em}\\
    $k$ $\mbox{SI}$ (nonlocal normal moment prior) &
    $\surd ([2 \lw{0}\{(1 + \sigma_\theta^2/\sigma^2)^{3/2}/(2 k e^{-1/2})\} - 1]\{1 + \sigma^2/\sigma_\theta^2\})$ \vspace{0.2em}\\
    $k$ $\mbox{minSI}$ (all priors) & $\surd (-2\log k)$ \vspace{0.2em}\\
    $k$ $\mbox{minSI}$ (local normal priors) & $\surd \{-\lw{-1}(-k^2/ e)\}$ \vspace{0.2em}\\
    $k$ $\mbox{minSI}$ ($-e\, p\, \log p$) &
    $\Phi^{-1}[1 - \exp\{\lw{-1}(-k/e)\}/2]$ \vspace{0.2em}\\
    \bottomrule
    \end{tabular}
  \end{table*}

Which type of support interval should data analysts use in practice? We believe
that the support interval based on a normal prior distribution is the most
intuitive for encoding external knowledge. This type should therefore be
preferably used whenever external knowledge is available. At the same time, the
support interval based on a local normal prior with unit-information variance
\citep{Kass1995b} seems to be a reasonable ``default'' choice in cases where no
external knowledge is available. Finally, we believe that minimum support
intervals are mostly useful for giving confidence intervals an evidential
interpretation due to the one-to-one mapping between the two.

It is also not clear which support level $k$ should be used for computing
support intervals. If space permits, we recommend to visualize the Bayes factor
as a function of the null value as in Figure~\ref{fig:example}. A similar approach
has also been proposed by \citet{Grunwald2023} under the name of E-posterior.
The Bayes factor visualization
provides readers with a more gradual assessment of support, and any desired $k$
support interval can be read off from it. If there are space constraints, a
compromise is to report support intervals for different levels (\eg{}
$k \in \{1/10, 1, 10\}$) or to present a forest plot with ``telescope'' style
support intervals with ascending support levels stacked on top of each other, as
in Figure~\ref{fig:comparison}. We are hesitant to recommend a ``default''
support level because any classification of support is arbitrary, just like the
95\% confidence level convention. We believe that $k = 1$ is perhaps the least
arbitrary default level, as it represents the tipping point at which the
included parameter values begin to receive support from the data (although not
necessarily strong support).

Other approaches for reinterpreting confidence intervals have been proposed. For
instance, \citet{Rafi2020} propose to rename confidence intervals to
``compatibility'' intervals and give their confidence level an information
theoretic interpretation. For example, a 95\%~confidence interval contains
parameter values with at most 4.3 bits refutational ``surprisal''. This notion
of compatibility is logically weaker than the notion of support considered in
this paper as a failure to refute a parameter value cannot establish that this
parameter value is supported without reference to alternatives
\citep{Greenland2022}. Compatibility intervals are in this sense similar to
minimum support intervals; without a specified prior under the alternative
hypothesis only the maximum surprisal/evidence \emph{against} the included
parameter values can be quantified.

We also showed how the coverage of $k$ support intervals with $k < 1$ is bounded
by $(1 - k)100\%$, which holds even under sequential analyses with optional
stopping. For instance, a $k = 1/20$ support interval has valid $95\%$ coverage.
Of course, such error rate guarantees rest on the assumption that the data model
has been correctly specified, which in most real world applications will be
violated to some extent. We do not see this as a problem for the evidential
interpretation of support intervals, which is usually of more concern to data
analysts. Evidential inference does not rely on a statistical model being
``true'' in some abstract sense. Bayes factors and support intervals simply
quantify the relative predictive performance that the combination of data model
and parameter distribution yield on out-of-sample data \citep{Kass1995,
  OHagan2004, Gneiting2007a, Fong2020}. Such ``descriptive inferential statistics'' are
especially important for the analysis of convenience data samples which
typically violate assumptions of the underlying statistical model
% , so that overconfident error-rate statements are not appropriate leading to
% overconfident error rate statements from $p$-values and confidence intervals
% are overconfident
\citep{Amrhein2019, Shafer2021}. In fact, even one of the best known proponents
of $p$-values --- R.A. Fisher --- noted ``For all purposes, and more
particularly for the \emph{communication} of the relevant evidence supplied by a
body of data, the values of the Mathematical Likelihood are better fitted to
analyse, summarize, and communicate statistical evidence of types too weak to
supply true probability statements'' \citep[p. 70]{Fisher1956} clearly
recognizing the importance of inferential tools based on relative likelihood for
making sense out of data.

\section*{Software and data}
The point estimate and 95\% confidence interval of the adjusted log hazard ratio
were extracted from the abstract of \citet{RECOVERY2021}. All analyses were
conducted in the R programming language version
4.3.0 \citep{R}. Code and data
for reproducing the results in this manuscript are available at
%\url{https://github.com/XXXX}.
\url{https://github.com/SamCH93/ECoCI}. A snapshot of the GitHub repository at
the time of writing this article is archived at
\url{https://doi.org/10.5281/zenodo.6723249}. An R package for calibration of
confidence intervals to (minimum) support intervals is available at
\url{https://CRAN.R-project.org/package=ciCalibrate}, see
Appendix~\ref{app:Rpkg} for an illustration.

\section*{Acknowledgments}
We thank Leonhard Held for helpful comments on an earlier version of the
manuscript. We thank Michael Lindon for interesting discussions and for letting
us know about his work on the connection between support and confidence sets. We
thank Glenn Shafer for attending us about R.A. Fisher's work on relative
likelihood. We thank Sander Greenland for valuable feedback on the first version
of the manuscript. We thank the editor Joshua Tebbs, the anonymous associate
editor and the anonymous reviewer for useful comments. Our acknowledgment of
these individuals does not imply their endorsement of this article. This work
was supported in part by an NWO Vici grant (016.Vici.170.083) to EJW, and a
Swiss National Science Foundation mobility grant (part of 189295) to SP.

%% Appendix
%% -----------------------------------------------------------------------------
\begin{appendices}

\section{The ciCalibrate package}
\label{app:Rpkg}
We provide an R implementation of the support intervals and underlying Bayes
factor functions from Table~\ref{tab:summary}. The package is available at
\url{https://CRAN.R-project.org/package=ciCalibrate} and can be installed by
executing \texttt{install.packages("ciCalibrate")} in an R console. The
following code snippet illustrates the computation and plotting of support
interval and Bayes factor function.

\begin{knitrout}
\definecolor{shadecolor}{rgb}{0.969, 0.969, 0.969}\color{fgcolor}\begin{kframe}
\begin{alltt}
\hlcom{## 95% CI from RECOVERY trial}
\hlstd{logHRci} \hlkwb{<-} \hlkwd{c}\hlstd{(}\hlopt{-}\hlnum{0.29}\hlstd{,} \hlopt{-}\hlnum{0.07}\hlstd{)}
\hlcom{## compute a support interval with level k = 10}
\hlkwd{library}\hlstd{(}\hlstr{"ciCalibrate"}\hlstd{)} \hlcom{# install with install.packages("ciCalibrate")}
\hlstd{si10} \hlkwb{<-} \hlkwd{ciCalibrate}\hlstd{(}\hlkwc{ci} \hlstd{= logHRci,} \hlkwc{ciLevel} \hlstd{=} \hlnum{0.95}\hlstd{,} \hlkwc{siLevel} \hlstd{=} \hlnum{10}\hlstd{,}
                    \hlkwc{method} \hlstd{=} \hlstr{"SI-normal"}\hlstd{,} \hlkwc{priorMean} \hlstd{=} \hlnum{0}\hlstd{,} \hlkwc{priorSD} \hlstd{=} \hlnum{2}\hlstd{)}
\hlstd{si10}
\end{alltt}
\begin{verbatim}
## 
## Point Estimate [95% Confidence Interval] 
## -0.18 [-0.29,-0.07]
## 
## Calibration Method
## Normal prior for parameter under alternative
## with mean m = 0 and standard deviation sd = 2
## 
## k = 10 Support Interval
## [-0.27,-0.09]
\end{verbatim}
\begin{alltt}
\hlcom{## plot Bayes factor function with support interval}
\hlkwd{plot}\hlstd{(si10)}
\end{alltt}
\end{kframe}
\includegraphics[width=\maxwidth]{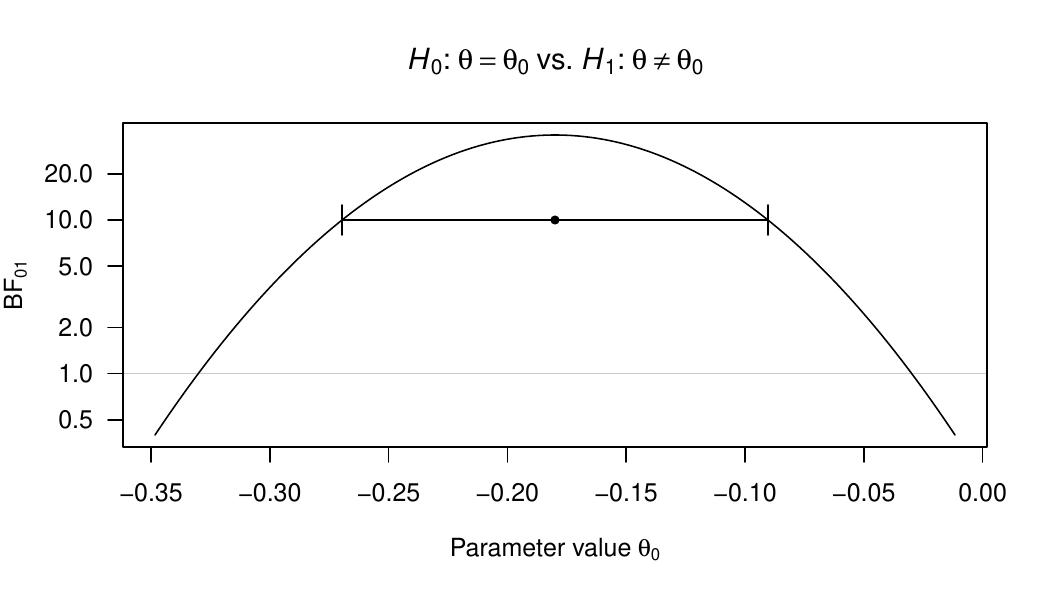} 
\end{knitrout}

\end{appendices}

\section*{Conflict of interest}
The authors report that there are no conflicts of interest to declare.

%% Bibliography
%% -----------------------------------------------------------------------------
\bibliographystyle{apalikedoiurl}
\bibliography{bibliography}

\end{document}

%% file: defs.tex
\newcommand{\eg}{e.\,g.,\,} % e.g.
\newcommand{\ie}{i.\,e.,\,} % i.e.
\DeclareMathOperator{\Nor}{N} % Normal
\renewcommand{\P}{\operatorname{\mbox{Pr}}} % Probability
\newcommand{\R}{\mathbb{R}} % Reals
\newcommand{\given}{\,\vert\,} % Given
\newcommand{\BF}{\mbox{BF}} % Bayes factor
\newcommand{\that}{\hat{\theta}} % Effect estimate
\newcommand{\minBF}{\mbox{minBF}} % minimum Bayes factor
\newcommand{\h}[1]{\mathcal{H}_{#1}} % Hypothesis
\newcommand{\lw}[1]{\mbox{W}_{\scriptscriptstyle #1}} % Lambert W function

%% file: ecoci.bbl
\begin{thebibliography}{}

\bibitem[Amrhein et~al., 2019]{Amrhein2019}
Amrhein, V., Trafimow, D., and Greenland, S. (2019).
\newblock Inferential statistics as descriptive statistics: There is no
  replication crisis if we don't expect replication.
\newblock {\em The American Statistician}, 73(sup1):262--270.
\newblock \doi{10.1080/00031305.2018.1543137}.

\bibitem[Berger et~al., 2013]{Berger2013}
Berger, J., Bayarri, M.~J., and Pericchi, L.~R. (2013).
\newblock The effective sample size.
\newblock {\em Econometric Reviews}, 33(1-4):197--217.
\newblock \doi{10.1080/07474938.2013.807157}.

\bibitem[Berger and Delampady, 1987]{Berger1987b}
Berger, J.~O. and Delampady, M. (1987).
\newblock Testing precise hypotheses.
\newblock {\em Statistical Science}, 2(3):317--335.
\newblock \doi{10.1214/ss/1177013238}.

\bibitem[Berger and Sellke, 1987]{Berger1987}
Berger, J.~O. and Sellke, T. (1987).
\newblock Testing a point null hypothesis: The irreconcilability of {$P$}
  values and evidence.
\newblock {\em Journal of the American Statistical Association}, 82(397):112.
\newblock \doi{10.2307/2289131}.

\bibitem[Blume, 2002]{Blume2002}
Blume, J.~D. (2002).
\newblock Likelihood methods for measuring statistical evidence.
\newblock {\em Statistics in Medicine}, 21(17):2563--2599.
\newblock \doi{10.1002/sim.1216}.

\bibitem[Corless et~al., 1996]{Corless1996}
Corless, R.~M., Gonnet, G.~H., Hare, D. E.~G., Jeffrey, D.~J., and Knuth, D.~E.
  (1996).
\newblock On the {Lambert W} function.
\newblock {\em Advances in Computational Mathematics}, 5(1):329--359.
\newblock \doi{10.1007/bf02124750}.

\bibitem[Edwards, 1971]{Edwards1971}
Edwards, A. W.~F. (1971).
\newblock {\em Likelihood}.
\newblock Cambridge University Press, London.

\bibitem[Edwards et~al., 1963]{Edwards1963}
Edwards, W., Lindman, H., and Savage, L.~J. (1963).
\newblock {Bayesian} statistical inference for psychological research.
\newblock {\em Psychological Review}, 70(3):193--242.
\newblock \doi{10.1037/h0044139}.

\bibitem[Fisher, 1956]{Fisher1956}
Fisher, R.~A. (1956).
\newblock {\em Statistical methods and scientific inference.}
\newblock Oliver \& Boyd, Edinburgh.

\bibitem[Fong and Holmes, 2020]{Fong2020}
Fong, E. and Holmes, C.~C. (2020).
\newblock On the marginal likelihood and cross-validation.
\newblock {\em Biometrika}, 107(2):489--496.
\newblock \doi{10.1093/biomet/asz077}.

\bibitem[Fraser, 2019]{Fraser2019}
Fraser, D. A.~S. (2019).
\newblock The \textit{p}-value function and statistical inference.
\newblock {\em The American Statistician}, 73(sup1):135--147.
\newblock \doi{10.1080/00031305.2018.1556735}.

\bibitem[Gneiting and Raftery, 2007]{Gneiting2007a}
Gneiting, T. and Raftery, E. (2007).
\newblock Strictly proper scoring rules, prediction, and estimation.
\newblock {\em Journal of the Amerian Statistical Association},
  102(477):359--377.
\newblock \doi{10.1198/016214506000001437}.

\bibitem[Good, 1992]{Good1992}
Good, I.~J. (1992).
\newblock The {Bayes}/non-{Bayes} compromise: A brief review.
\newblock {\em Journal of the American Statistical Association},
  87(419):597--606.
\newblock \doi{10.1080/01621459.1992.10475256}.

\bibitem[Greenland, 2023]{Greenland2022}
Greenland, S. (2023).
\newblock Divergence versus decision \textit{{P}}-values: A distinction worth
  making in theory and keeping in practice: Or, how divergence
  \textit{{P}}-values measure evidence even when decision \textit{{P}}-values
  do not.
\newblock {\em Scandinavian Journal of Statistics}, 50(1):54--88.
\newblock \doi{10.1111/sjos.12625}.

\bibitem[Greenland et~al., 2016]{Greenland2016}
Greenland, S., Senn, S.~J., Rothman, K.~J., Carlin, J.~B., Poole, C., Goodman,
  S.~N., and Altman, D.~G. (2016).
\newblock Statistical tests, \textit{P} values, confidence intervals, and
  power: a guide to misinterpretations.
\newblock {\em European Journal of Epidemiology}, 31(4):337--350.
\newblock \doi{10.1007/s10654-016-0149-3}.

\bibitem[Gr\"{u}nwald et~al., 2019]{Grundwald2019}
Gr\"{u}nwald, P., de~Heide, R., and Koolen, W. (2019).
\newblock Safe testing.
\newblock \doi{10.48550/ARXIV.1906.07801}.
\newblock Preprint.

\bibitem[Grünwald, 2023]{Grunwald2023}
Grünwald, P. (2023).
\newblock The {E}-posterior.
\newblock {\em Philosophical Transactions of the Royal Society A}, 381(2247).
\newblock \doi{10.1098/rsta.2022.0146}.

\bibitem[Hacking, 1965]{Hacking1965}
Hacking, I. (1965).
\newblock {\em Logic of Statistical Inference}.
\newblock Cambridge University Press, New York.

\bibitem[Held and Ott, 2018]{Held2018}
Held, L. and Ott, M. (2018).
\newblock On \textit{p}-values and {Bayes} factors.
\newblock {\em Annual Review of Statistics and Its Application}, 5(1):393--419.
\newblock \doi{10.1146/annurev-statistics-031017-100307}.

\bibitem[Hendriksen et~al., 2021]{Hendriksen2021}
Hendriksen, A., de~Heide, R., and Gr\"{u}nwald, P. (2021).
\newblock Optional stopping with {Bayes} factors: A categorization and
  extension of folklore results, with an application to invariant situations.
\newblock {\em Bayesian Analysis}, 16(3):961--989.
\newblock \doi{10.1214/20-ba1234}.

\bibitem[Hoekstra et~al., 2014]{Hoekstra2014}
Hoekstra, R., Morey, R.~D., Rouder, J.~N., and Wagenmakers, E.-J. (2014).
\newblock Robust misinterpretation of confidence intervals.
\newblock {\em Psychonomic Bulletin \& Review volume}, 21(5):1157--1164.
\newblock \doi{10.3758/s13423-013-0572-3}.

\bibitem[Howard et~al., 2021]{Howard2021}
Howard, S.~R., Ramdas, A., McAuliffe, J., and Sekhon, J. (2021).
\newblock Time-uniform, nonparametric, nonasymptotic confidence sequences.
\newblock {\em The Annals of Statistics}, 49(2):1055--1080.
\newblock \doi{10.1214/20-aos1991}.

\bibitem[Jeffreys, 1961]{Jeffreys1961}
Jeffreys, H. (1961).
\newblock {\em Theory of Probability}.
\newblock Oxford: Clarendon Press, third edition.

\bibitem[Johnson and Rossell, 2010]{Johnson2010}
Johnson, V.~E. and Rossell, D. (2010).
\newblock On the use of non-local prior densities in {Bayesian} hypothesis
  tests.
\newblock {\em Journal of the Royal Statistical Society: Series B (Statistical
  Methodology)}, 72(2):143--170.
\newblock \doi{10.1111/j.1467-9868.2009.00730.x}.

\bibitem[Kass and Raftery, 1995]{Kass1995}
Kass, R.~E. and Raftery, A.~E. (1995).
\newblock Bayes factors.
\newblock {\em Journal of the American Statistical Association},
  90(430):773--795.
\newblock \doi{10.1080/01621459.1995.10476572}.

\bibitem[Kass and Wasserman, 1995]{Kass1995b}
Kass, R.~E. and Wasserman, L. (1995).
\newblock A reference {Bayesian} test for nested hypotheses and its
  relationship to the {Schwarz} criterion.
\newblock {\em Journal of the American Statistical Association},
  90(431):928--934.
\newblock \doi{10.1080/01621459.1995.10476592}.

\bibitem[Lai, 1976]{Lai1976}
Lai, T.~L. (1976).
\newblock On confidence sequences.
\newblock {\em The Annals of Statistics}, 4(2).
\newblock \doi{10.1214/aos/1176343406}.

\bibitem[Lindon and Malek, 2020]{Lindon2020}
Lindon, M. and Malek, A. (2020).
\newblock Sequential testing of multinomial hypotheses with applications to
  detecting implementation errors and missing data in randomized experiments.
\newblock URL \url{https://arxiv.org/abs/2011.03567v1}.

\bibitem[Ly et~al., 2017]{Ly2017}
Ly, A., Marsman, M., Verhagen, J., Grasman, R.~P., and Wagenmakers, E.-J.
  (2017).
\newblock A tutorial on {Fisher} information.
\newblock {\em Journal of Mathematical Psychology}, 80:40--55.
\newblock \doi{10.1016/j.jmp.2017.05.006}.

\bibitem[O'Hagan and Forster, 2004]{OHagan2004}
O'Hagan, A. and Forster, J.~J. (2004).
\newblock {\em Kendall's Advanced Theory of Statistics, volume 2B: {Bayesian}
  Inference}.
\newblock Arnold, London, UK, second edition.

\bibitem[Pace and Salvan, 2020]{Pace2020}
Pace, L. and Salvan, A. (2020).
\newblock Likelihood, replicability and {Robbins}{\textquotesingle} confidence
  sequences.
\newblock {\em International Statistical Review}, 88(3):599--615.
\newblock \doi{10.1111/insr.12355}.

\bibitem[Pramanik and Johnson, 2022]{Pramanik2022}
Pramanik, S. and Johnson, V.~E. (2022).
\newblock Efficient alternatives for {Bayesian} hypothesis tests in psychology.
\newblock {\em Psychological Methods}.
\newblock \doi{10.1037/met0000482}.

\bibitem[{R Core Team}, 2023]{R}
{R Core Team} (2023).
\newblock {\em R: A Language and Environment for Statistical Computing}.
\newblock R Foundation for Statistical Computing, Vienna, Austria.
\newblock URL \url{https://www.R-project.org/}.

\bibitem[Rafi and Greenland, 2020]{Rafi2020}
Rafi, Z. and Greenland, S. (2020).
\newblock Semantic and cognitive tools to aid statistical science: replace
  confidence and significance by compatibility and surprise.
\newblock {\em BMC Medical Research Methodology}, 20(1):244.
\newblock \doi{10.1186/s12874-020-01105-9}.

\bibitem[Raftery, 1999]{Raftery1999}
Raftery, A.~E. (1999).
\newblock Bayes factors and {BIC}.
\newblock {\em Sociological Methods \& Research}, 27(3):411--427.
\newblock \doi{10.1177/0049124199027003005}.

\bibitem[{RECOVERY Collaborative Group}, 2021]{RECOVERY2021}
{RECOVERY Collaborative Group} (2021).
\newblock Dexamethasone in hospitalized patients with {Covid}-19.
\newblock {\em New England Journal of Medicine}, 384(8):693--704.
\newblock \doi{10.1056/nejmoa2021436}.

\bibitem[Robbins, 1970]{Robbins1970}
Robbins, H. (1970).
\newblock Statistical methods related to the law of the iterated logarithm.
\newblock {\em The Annals of Mathematical Statistics}, 41(5):1397--1409.
\newblock \doi{10.1214/aoms/1177696786}.

\bibitem[Royall, 1997]{Royall1997}
Royall, R. (1997).
\newblock {\em Statistical evidence: a likelihood paradigm}.
\newblock Chapman \& Hall, London New York.

\bibitem[Sellke et~al., 2001]{Selke2001}
Sellke, T., Bayarri, M.~J., and Berger, J.~O. (2001).
\newblock Calibration of \textit{p} values for testing precise null hypotheses.
\newblock 55(1):62--71.
\newblock \doi{10.1198/000313001300339950}.

\bibitem[Shafer, 2021]{Shafer2021}
Shafer, G. (2021).
\newblock Descriptive probability.
\newblock Working paper \#59 (version September 30, 2021).
  \url{http://probabilityandfinance.com/articles/59.pdf}.

\bibitem[Spiegelhalter et~al., 2004]{Spiegelhalter2004}
Spiegelhalter, D.~J., Abrams, R., and Myles, J.~P. (2004).
\newblock {\em {Bayesian} Approaches to Clinical Trials and Health-Care
  Evaluation}.
\newblock New York: Wiley.

\bibitem[Vovk, 1993]{Vovk1993}
Vovk, V.~G. (1993).
\newblock A logic of probability, with application to the foundations of
  statistics.
\newblock {\em Journal of the Royal Statistical Society: Series B
  (Methodological)}, 55(2):317--341.
\newblock \doi{10.1111/j.2517-6161.1993.tb01904.x}.

\bibitem[Wagenmakers, 2022]{Wagenmakers2022}
Wagenmakers, E.-J. (2022).
\newblock Approximate objective {Bayes} factors from \textit{P}-values and
  sample size: The $3p\sqrt{n}$ rule.
\newblock \doi{10.31234/osf.io/egydq}.

\bibitem[Wagenmakers et~al., 2022]{Wagenmakers2020}
Wagenmakers, E.-J., Gronau, Q.~F., Dablander, F., and Etz, A. (2022).
\newblock The support interval.
\newblock {\em Erkenntnis}, 87:589--601.
\newblock \doi{10.1007/s10670-019-00209-z}.

\bibitem[Wagenmakers and Ly, 2023]{Wagenmakers2021a}
Wagenmakers, E.-J. and Ly, A. (2023).
\newblock History and nature of the {Jeffreys-Lindley} paradox.
\newblock {\em Archive for History of Exact Sciences}, 77:25--72.
\newblock \doi{10.1007/s00407-022-00298-3}.

\bibitem[Wassmer and Brannath, 2016]{Wassmer2016}
Wassmer, G. and Brannath, W. (2016).
\newblock {\em Group Sequential and Confirmatory Adaptive Designs in Clinical
  Trials}.
\newblock Springer International Publishing, Cham, Switzerland.
\newblock \doi{10.1007/978-3-319-32562-0}.

\end{thebibliography}
